\documentclass[final]{aipproc}
\layoutstyle{6x9}

\usepackage{float}
\usepackage{floatflt}

\usepackage[centertags]{amsmath}
\allowdisplaybreaks[1]

\begin{document}

\title{Double Beta Decay and the Absolute\\Neutrino Mass Scale\footnote{
Talk presennted at
NuFact 03, 5$^{\mathrm{th}}$ International Workshop on Neutrino Factories \& Superbeams,
5--11 June 2003, Columbia University, New York.}}

\author{Carlo Giunti}{
address={INFN, Sezione di Torino, and Dipartimento di Fisica Teorica,
\\
Universit\`a di Torino,
Via P. Giuria 1, I--10125 Torino, Italy}
}

\begin{abstract}
After a short review of the current status of three-neutrino mixing,
the implications for the values of neutrino masses are discussed.
The bounds on the absolute scale of neutrino masses from Tritium $\beta$-decay
and cosmological data are reviewed.
Finally, we discuss
the implications of three-neutrino mixing for neutrinoless
double-$\beta$ decay.
\end{abstract}

\maketitle



The marvelous recent results of
neutrino oscillation experiments
have given us important information on neutrino mixing.
The data of solar neutrino experiments
and of the KamLAND long-baseline $\bar\nu_e$ disappearance experiment
show $\nu_e \to \nu_\mu, \nu_\tau$
transitions generated by
the squared-mass difference
$\Delta{m}^2_{\mathrm{SUN}}$
in one of the two ranges
\cite{hep-ph/0212129}
\begin{equation}
5.1 \times 10^{-5}
<
\Delta{m}^2_{\mathrm{SUN}}
<
9.7 \times 10^{-5}
(\text{\footnotesize{LMA-I}})
\,,
\
1.2 \times 10^{-4}
<
\Delta{m}^2_{\mathrm{SUN}}
<
1.9 \times 10^{-4}
(\text{\footnotesize{LMA-II}})
\,,
\label{001}
\end{equation}
at 99.73\% C.L.,
with best-fit value
$ \Delta{m}^{2\,\mathrm{bf}}_{\mathrm{SUN}} \simeq 6.9 \times 10^{-5} $
(we measure squared-mass differences in units of eV$^2$).
The effective solar mixing angle $\vartheta_{\mathrm{SUN}}$
is constrained at 99.73\% C.L. in the interval
\cite{hep-ph/0212129}
\begin{equation}
0.29 < \tan^2 \vartheta_{\mathrm{SUN}} < 0.86
\,,
\label{002}
\end{equation}
with best-fit value
$ \tan^2 \vartheta_{\mathrm{SUN}}^{\mathrm{bf}} \simeq 0.46 $.
The results of atmospheric neutrino experiments
and of the K2K long-baseline $\nu_\mu$ disappearance experiment
indicate
$\nu_\mu \to \nu_\tau$
transitions generated by
the squared-mass difference
$\Delta{m}^2_{\mathrm{ATM}}$
in the 99.73\% C.L. range
\cite{hep-ph/0303064}
\begin{equation}
1.4 \times 10^{-3}
<
\Delta{m}^2_{\mathrm{ATM}}
<
5.1 \times 10^{-3}
\,,
\label{003}
\end{equation}
with best-fit value
$
\Delta{m}^{2\,\mathrm{bf}}_{\mathrm{ATM}}
\simeq
2.6 \times 10^{-3}
$.
The best-fit effective atmospheric mixing angle $\vartheta_{\mathrm{ATM}}$
is maximal,
$
\sin^2 2 \vartheta_{\mathrm{ATM}}^{\mathrm{bf}} \simeq 1
$,
with the 99.73\% C.L. lower bound
\cite{hep-ph/0303064}
\begin{equation}
\sin^2 2 \vartheta_{\mathrm{ATM}} > 0.86
\,.
\label{004}
\end{equation}
These evidences of neutrino mixing
are nicely accommodated in the minimal framework of three-neutrino mixing,
in which the three flavor neutrinos
$\nu_e$,
$\nu_\mu$,
$\nu_\tau$
are unitary linear combinations of
three neutrinos
$\nu_1$,
$\nu_2$,
$\nu_3$
with masses
$m_1$,
$m_2$,
$m_3$,
respectively
(see Ref.~\cite{BGG-review-98}).
Figure~\ref{3nu}
shows the two three-neutrino schemes
allowed by the observed hierarchy
$\Delta{m}^2_{\mathrm{SUN}} \ll \Delta{m}^2_{\mathrm{ATM}}$,
with
the massive neutrinos labeled in order to have
\begin{equation}
\Delta{m}^2_{\mathrm{SUN}}
=
\Delta{m}^2_{21}
\,,
\qquad
\Delta{m}^2_{\mathrm{ATM}}
\simeq
|\Delta{m}^2_{31}|
\simeq
|\Delta{m}^2_{32}|
\,.
\label{006}
\end{equation}
The two schemes
in Fig.~\ref{3nu} are usually called
``normal''
and
``inverted'',
because in the normal scheme the smallest
squared-mass difference is generated by the two lightest neutrinos
and a natural neutrino mass hierarchy can be realized if
$m_1 \ll m_2$,
whereas in the inverted scheme the smallest
squared-mass difference is generated
by the two heaviest neutrinos,
which are almost degenerate for any value of the lightest neutrino mass $m_3$.
\begin{floatingfigure}[l]{0.43\textwidth}
\begin{flushleft}
\begin{minipage}[t]{0.43\textwidth}
\setlength{\tabcolsep}{0cm}
\begin{tabular}{lr}
\includegraphics*[bb=181 466 428 775, width=0.49\textwidth]{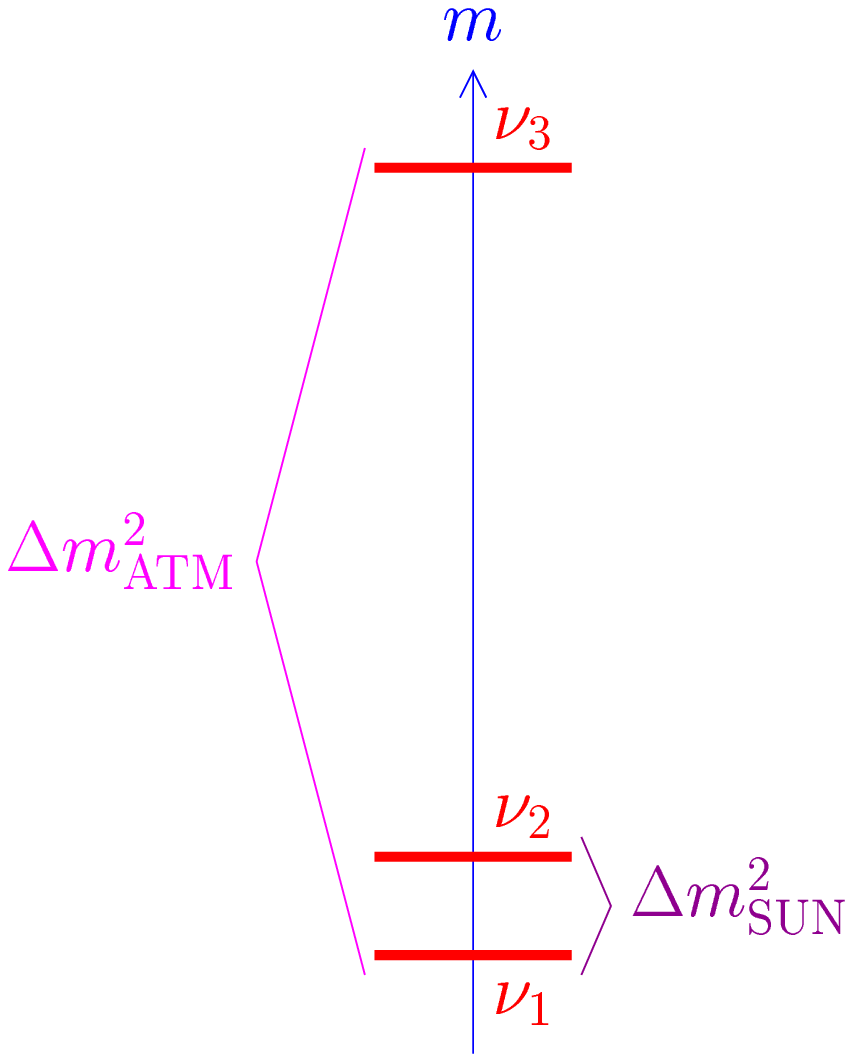}
&
\includegraphics*[bb=183 466 432 775, width=0.49\textwidth]{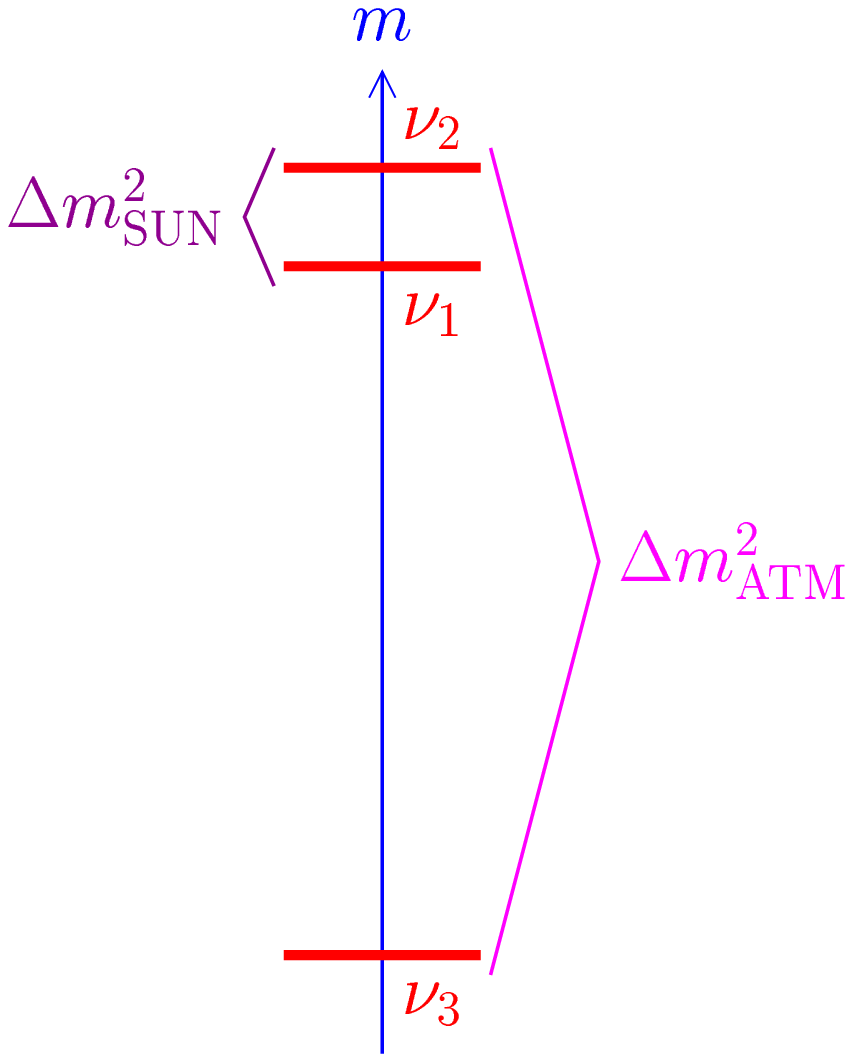}
\\
\textsf{normal}
&
\textsf{inverted}
\end{tabular}
\caption{ \label{3nu}
The two three-neutrino schemes allowed by the observed hierarchy
$\Delta{m}^2_{\mathrm{SUN}} \ll \Delta{m}^2_{\mathrm{ATM}}$.
}
\end{minipage}
\end{flushleft}
\end{floatingfigure}
\begin{figure}[b]
\begin{minipage}[t]{0.47\textwidth}
\begin{center}
\includegraphics*[bb=118 377 465 702, width=0.99\textwidth]{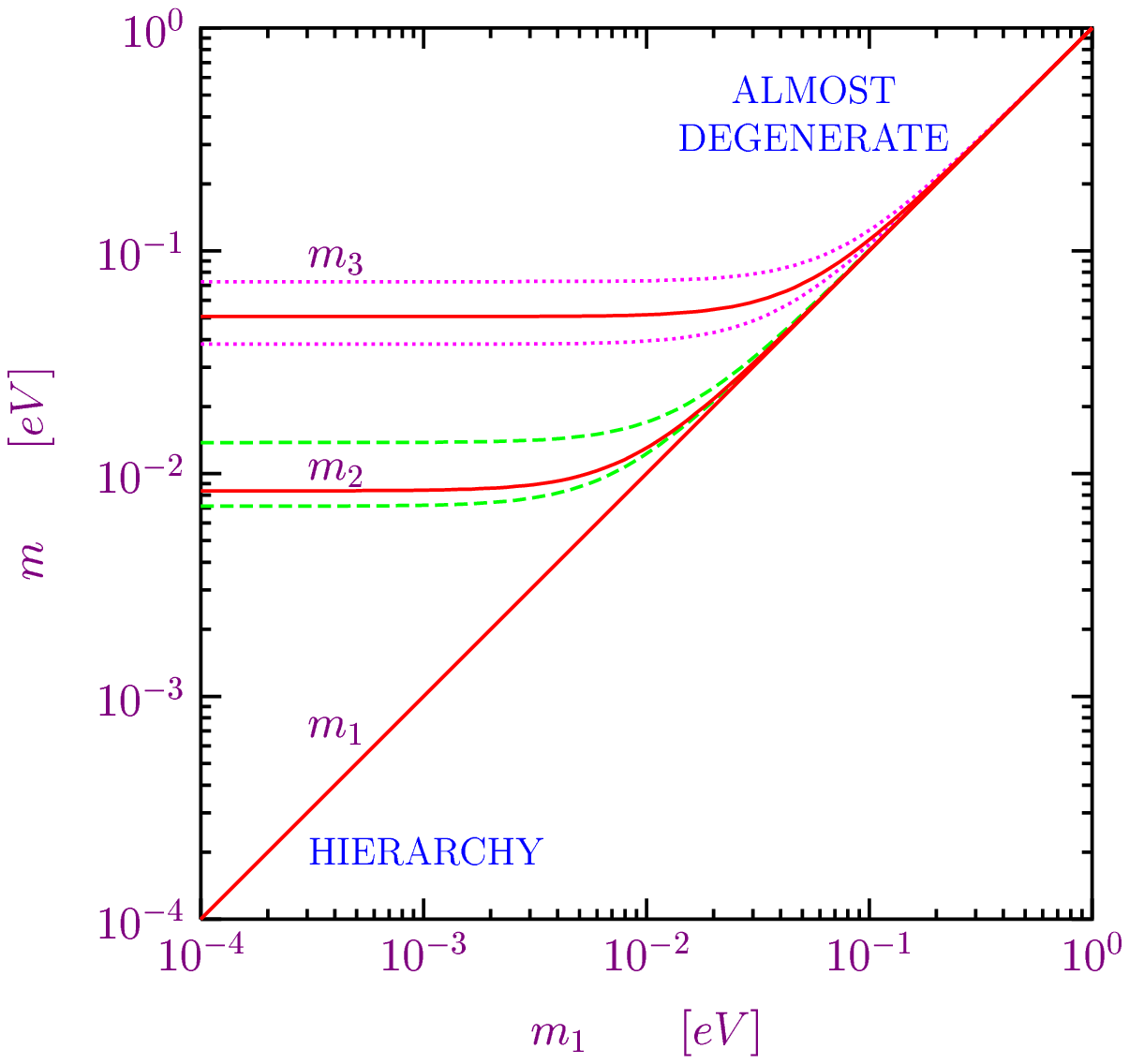}
\end{center}
\end{minipage}
\hfill
\begin{minipage}[t]{0.47\textwidth}
\begin{center}
\includegraphics*[bb=118 427 465 753, width=0.99\textwidth]{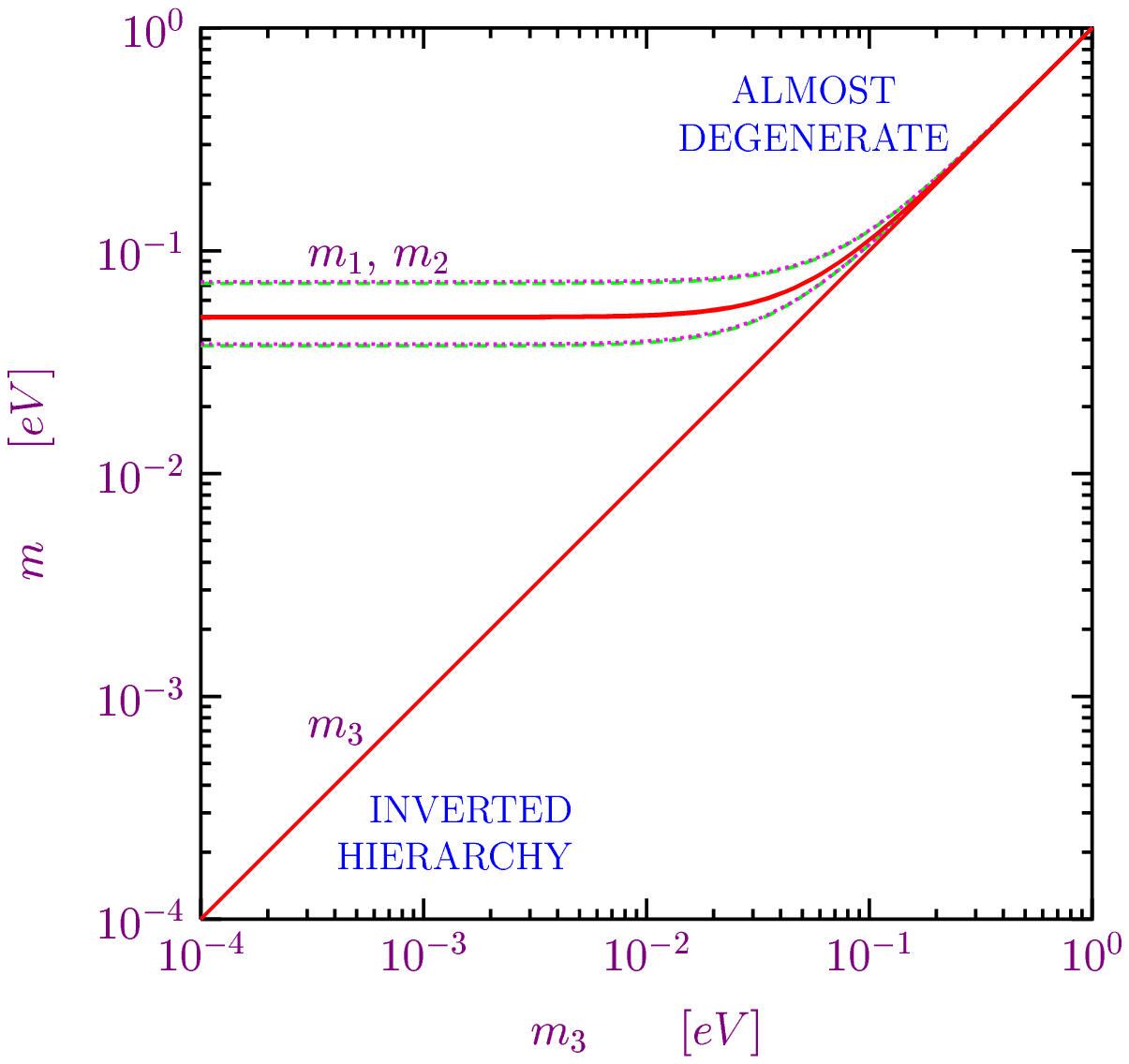}
\end{center}
\end{minipage}
\caption{ \label{3ma}
Allowed ranges for the neutrino masses as functions
of the lightest mass $m_1$ and $m_3$ in the normal and inverted
three-neutrino scheme, respectively.
}
\end{figure}
Solar neutrino oscillations depend only on the first row
$U_{e1}$,
$U_{e2}$,
$U_{e3}$
of the mixing matrix,
and
the hierarchy $\Delta{m}^2_{\mathrm{SUN}} \ll \Delta{m}^2_{\mathrm{ATM}}$
implies that neutrino oscillations generated by
$\Delta{m}^2_{\mathrm{ATM}}$
depend only on the last column
$U_{e3}$,
$U_{\mu3}$,
$U_{\tau3}$
of the mixing matrix.
Hence,
the only connection between
solar and atmospheric neutrino oscillations
is due to the element
$U_{e3}$.
The negative result of the CHOOZ long-baseline
$\bar\nu_e$ disappearance experiment
implies that electron neutrinos do not oscillate
at the atmospheric scale and
$|U_{e3}|$
is small:
$ |U_{e3}|^2 < 5 \times 10^{-2} $
at 99.73\% C.L.
\cite{Fogli:2002pb}.
Therefore,
solar and atmospheric neutrino oscillations are practically decoupled
\cite{Bilenky:1998tw}
and the effective mixing angles in
solar, atmospheric and CHOOZ experiments
can be related to the elements of the three-neutrino mixing matrix by
(see also Ref.~\cite{hep-ph/0212142})
\begin{equation}
\sin^2\vartheta_{\mathrm{SUN}}
=
\frac{|U_{e2}|^2}{1-|U_{e3}|^2}
\qquad
\sin^2\vartheta_{\mathrm{ATM}}
=
|U_{\mu3}|^2
\qquad
\sin^2\vartheta_{\mathrm{CHOOZ}}
=
|U_{e3}|^2
\,.
\label{011}
\end{equation}
Taking into account all the above experimental constraints,
we have reconstructed the best-fit and allowed ranges for the
elements of the mixing matrix
(see Ref.~\cite{hep-ph/0306001}
for a reconstruction
taking into account the correlations
among the mixing parameters):
\begin{equation}
U_{\mathrm{bf}}
\simeq
\left( \begin{smallmatrix}
-0.83 & 0.56 &  0.00 \\
 0.40 & 0.59 &  0.71 \\
 0.40 & 0.59 & -0.71
\end{smallmatrix} \right)
\,,
\quad
|U|
\simeq
\left( \begin{smallmatrix}
0.71-0.88 & 0.46-0.68 & 0.00-0.22 \\
0.08-0.66 & 0.26-0.79 & 0.55-0.85 \\
0.10-0.66 & 0.28-0.80 & 0.51-0.83
\end{smallmatrix} \right)
\,.
\label{012}
\end{equation}
Such mixing matrix,
with all elements large except $U_{e3}$,
is called ``bilarge''.
It is very different from the quark mixing matrix.

The absolute scale of neutrino masses
is not determined by the observation of
neutrino oscillations,
which
depend
only on the differences of the squares of neutrino masses.
Figure~\ref{3ma} shows the allowed ranges
(between the dashed and dotted curves)
for the neutrino masses
obtained from the allowed values of the oscillation parameters
in Eqs.~(\ref{001})--(\ref{004}),
as functions of the lightest mass
in the normal and inverted three-neutrino schemes.
The solid lines correspond to the best fit values of the oscillation parameters.
One can see that at least two neutrinos have masses
larger than about
$7 \times 10^{-3} \, \mathrm{eV}$.

The most sensitive known ways to probe the
absolute values of neutrino masses
are
the observation of the end-point part of
the electron spectrum in Tritium $\beta$-decay,
the observation of large-scale structures
in the early universe
and
the search for neutrinoless double-$\beta$ decay,
if neutrinos are Majorana particles
(see Ref.~\cite{Bilenky:2002aw};
we do not consider here the interesting possibility
to determine neutrino masses through the
observation of supernova neutrinos).


\begin{figure}[t]
\begin{minipage}[t]{0.47\textwidth}
\begin{center}
\includegraphics*[bb=121 376 465 702, width=0.99\textwidth]{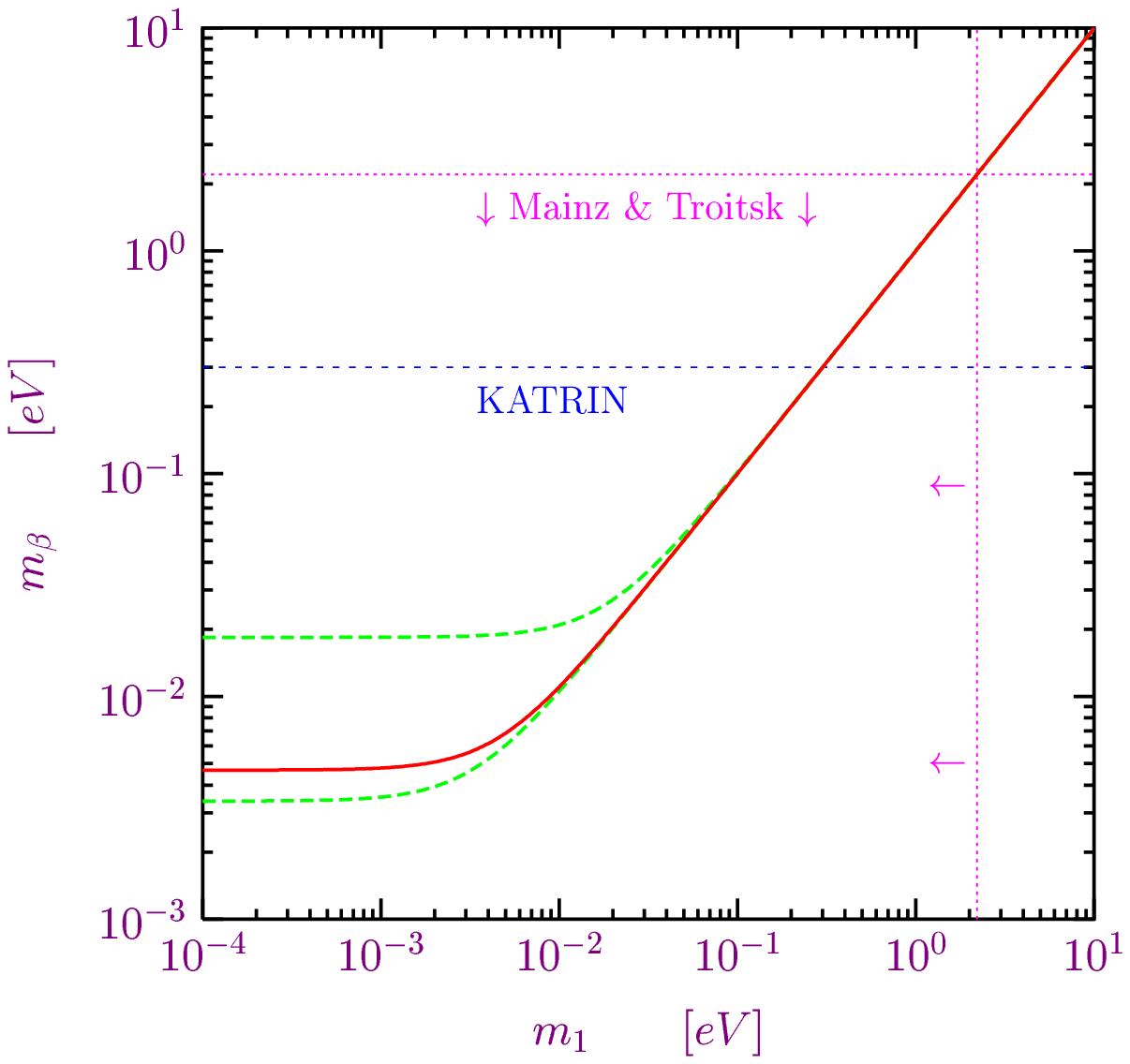}
\end{center}
\end{minipage}
\hfill
\begin{minipage}[t]{0.47\textwidth}
\begin{center}
\includegraphics*[bb=121 376 465 702, width=0.99\textwidth]{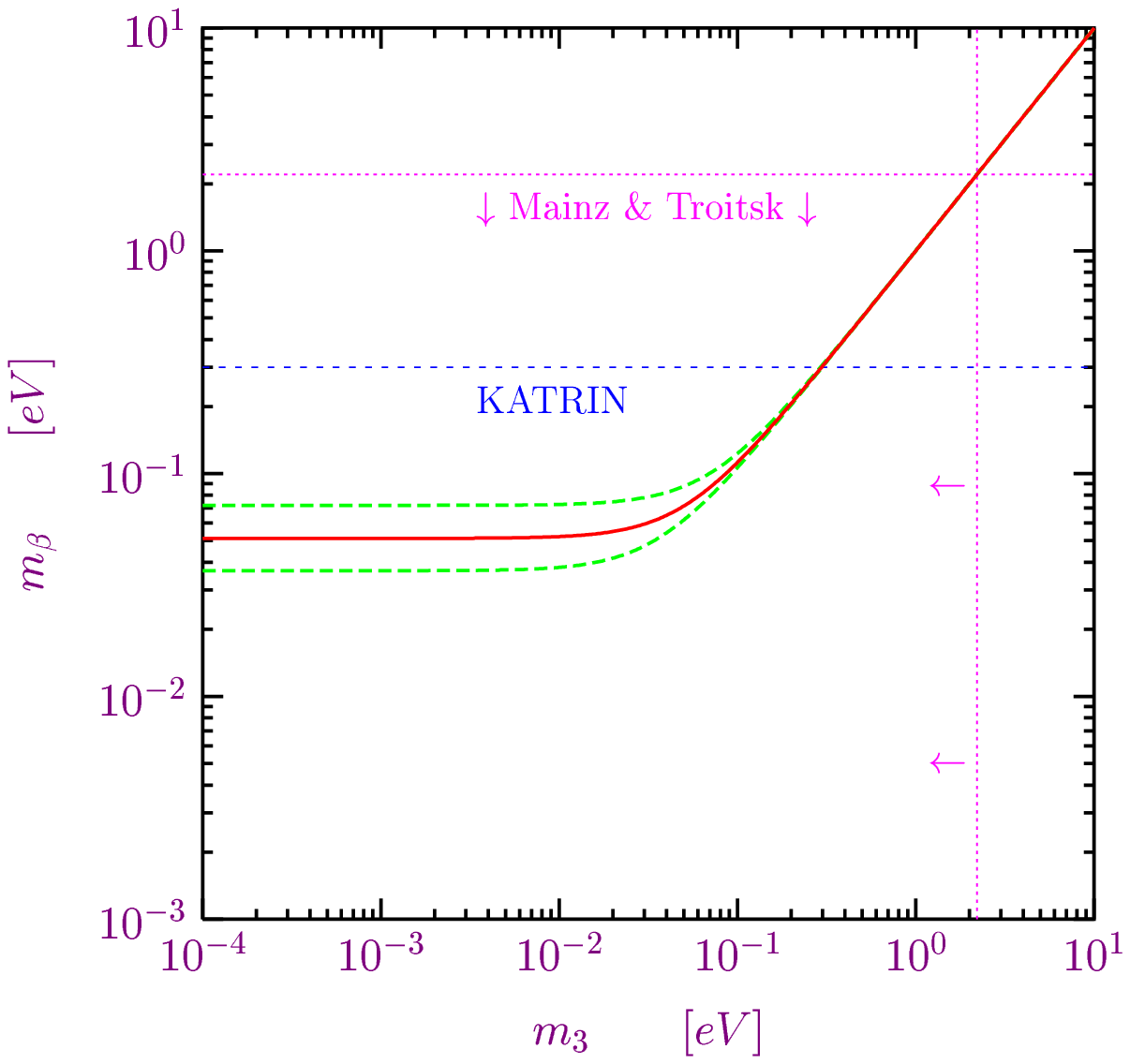}
\end{center}
\end{minipage}
\caption{ \label{mb}
Effective neutrino mass $m_\beta$
in Tritium $\beta$-decay experiments as a function
of the lightest mass $m_1$ and $m_3$ in the normal and inverted
three-neutrino scheme, respectively.
}
\end{figure}

Up to now,
no indication of a neutrino mass has been found in
Tritium $\beta$-decay experiments,
leading to an upper limit on the effective mass
\begin{equation}
m_\beta = \sqrt{ \sum_k |U_{ek}|^2 m_k^2 }
\label{007}
\end{equation}
of $2.2 \, \mathrm{eV}$
at 95\% C.L.
\cite{hep-ex/0210050},
obtained in the Mainz and Troitsk experiments.
After 2007, the KATRIN experiment
\cite{hep-ex/0109033}
will explore $m_\beta$ down to about
$0.2-0.3 \, \mathrm{eV}$.
Figure~\ref{mb} shows the allowed range (between the dashed curves)
for $m_\beta$
obtained from the allowed values of the oscillation parameters
in Eqs.~(\ref{001})--(\ref{004}),
as a function of the lightest mass
in the normal and inverted three-neutrino schemes.
The solid line corresponds to the best fit values of the oscillation parameters.
One can see that in the normal scheme with a mass hierarchy
$m_\beta$
has a value between about
$3 \times 10^{-3} \, \mathrm{eV}$
and
$2 \times 10^{-2} \, \mathrm{eV}$,
whereas in the inverted scheme
$m_\beta$
is larger than about
$3 \times 10^{-2} \, \mathrm{eV}$.
Therefore,
if in the future it will be possible to constraint
$m_\beta$
to be smaller than about
$3 \times 10^{-2} \, \mathrm{eV}$,
a normal hierarchy of neutrino masses will be established.


The analysis of recent data on cosmic microwave background radiation
and
large scale structure in the universe
in the framework of the standard cosmological model
has allowed to establish an upper bound of
about 1 eV for the sum of neutrino masses,
which implies an upper limit of about 0.3 eV
for the individual masses
\cite{Spergel:2003cb,astro-ph/0303076}.
This limit is already at the same level as the sensitivity of the
future KATRIN experiment.
Let us emphasize,
however,
that the KATRIN experiment is important in order to probe
the neutrino masses in a model-independent way,


\begin{figure}[t]
\begin{minipage}[t]{0.47\textwidth}
\begin{center}
\includegraphics*[bb=121 376 465 702, width=0.99\textwidth]{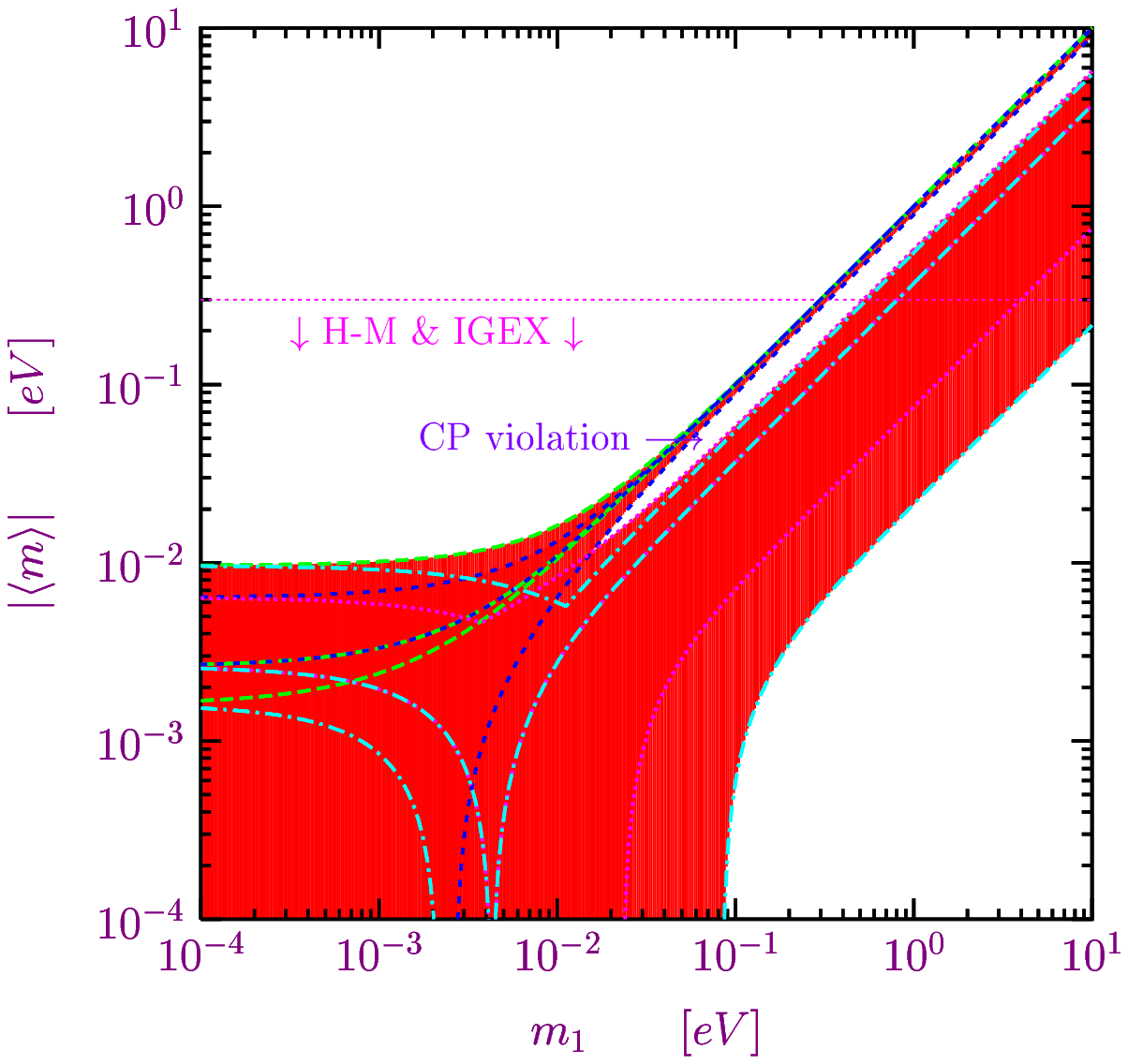}
\end{center}
\end{minipage}
\hfill
\begin{minipage}[t]{0.47\textwidth}
\begin{center}
\includegraphics*[bb=121 376 465 702, width=0.99\textwidth]{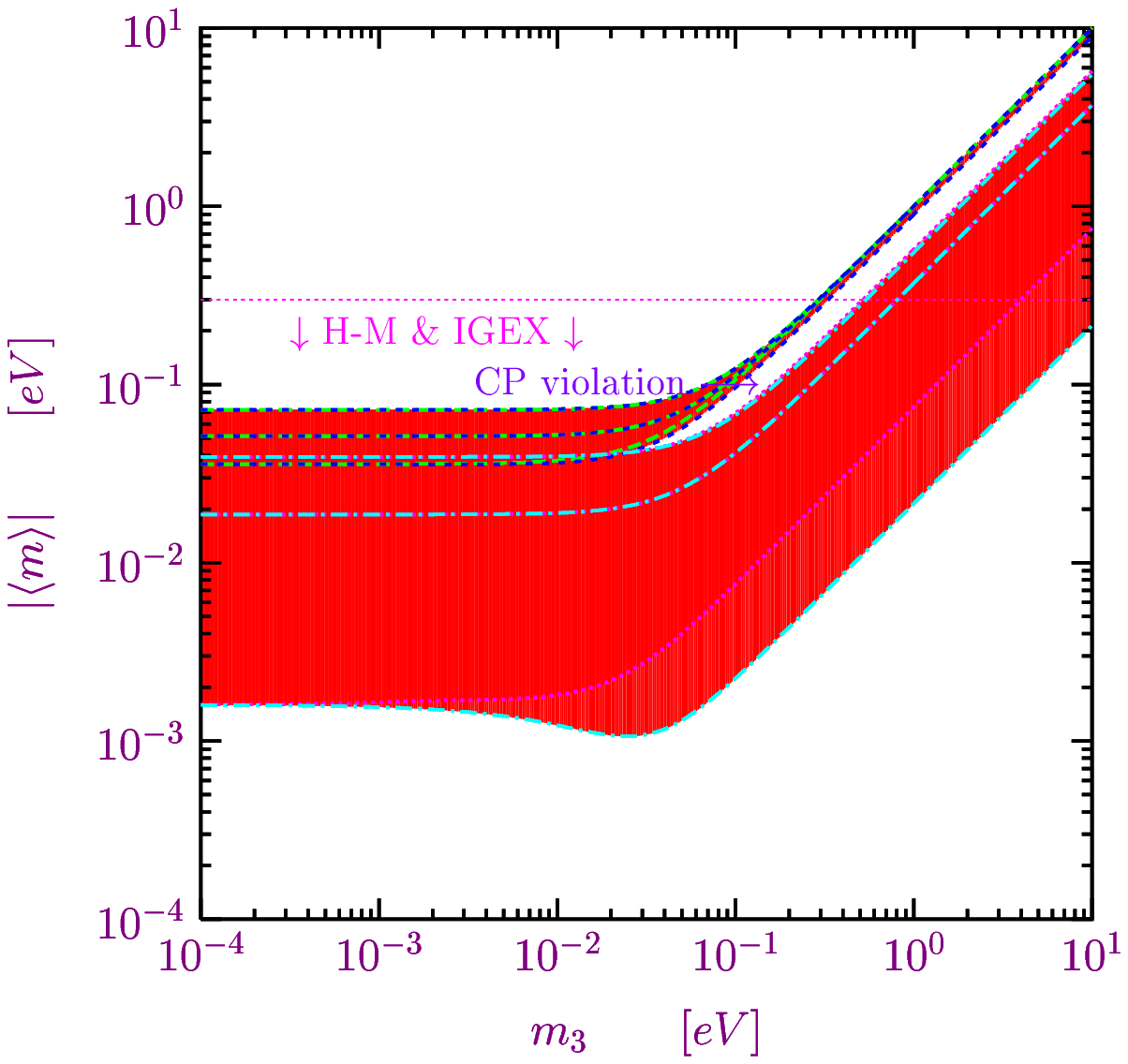}
\end{center}
\end{minipage}
\caption{ \label{db}
Effective Majorana mass $|\langle{m}\rangle|$
in neutrinoless double-$\beta$ decay experiments as a function
of the lightest mass $m_1$ and $m_3$ in the normal and inverted
three-neutrino scheme, respectively.
}
\end{figure}

A very important open problem in neutrino physics
is the Dirac or Majorana nature of neutrinos.
From the theoretical point of view it
is expected that neutrinos are Majorana particles,
with masses generated by effective Lagrangian terms
in which heavy degrees of freedom have been integrated out
(see Ref.~\cite{Altarelli:2003vk}).
In this case the smallness of neutrino masses
is naturally explained by the suppression due to the
ratio of the electroweak symmetry breaking scale
and
a high energy scale associated with the violation of the total lepton number
and new physics beyond the Standard Model.

The best known way to search for Majorana neutrino masses
is neutrinoless double-$\beta$ decay,
whose amplitude is proportional to the effective Majorana mass
\begin{equation}
|\langle{m}\rangle|
=
\bigg|
\sum_k
U_{ek}^2 \, m_k
\bigg|
\,.
\label{021}
\end{equation}
The present experimental upper limit on $|\langle{m}\rangle|$
between about 0.3 eV and 1.3 eV has been
obtained in the Heidelberg-Moscow and IGEX experiments.
The large uncertainty is due to the difficulty
of calculating the nuclear matrix element in the decay.
Figure~\ref{db} shows the allowed range for $|\langle{m}\rangle|$
obtained from the allowed values of the oscillation parameters
in Eqs.~(\ref{001})--(\ref{004}),
as a function of the lightest mass
in the normal and inverted three-neutrino schemes
(see also Ref.~\cite{Pascoli:2002qm}).
If CP is conserved,
$|\langle{m}\rangle|$
is constrained to lie in a shadowed region.
Finding $|\langle{m}\rangle|$ in an unshaded strip
would signal CP violation.
One can see that in the normal scheme large cancellations
between the three mass contributions are possible
and
$|\langle{m}\rangle|$
can be arbitrarily small.
On the other hand,
the cancellations in the inverted scheme are limited,
because $\nu_1$ and $\nu_2$,
with which the electron neutrino has large mixing,
are almost degenerate and much heavier than $\nu_3$.
Since the solar mixing angle is less than maximal,
a complete cancellation between the contributions of $\nu_1$ and $\nu_2$
is excluded,
leading to a lower bound of about
$1 \times 10^{-3} \, \mathrm{eV}$
for $|\langle{m}\rangle|$
in the inverted scheme.
If in the future
$|\langle{m}\rangle|$
will be found to be smaller than
about
$1 \times 10^{-3} \, \mathrm{eV}$,
it will be established that either neutrinos have a mass hierarchy
or they are Dirac particles.
Many neutrinoless double-$\beta$ decay experiments are planned for the future,
but they will unfortunately not be able to probe such small
values of $|\langle{m}\rangle|$,
extending their sensitivity at most in the
$10^{-2} \, \mathrm{eV}$ range
(see Ref.~\cite{Bilenky:2002aw}).

In conclusion,
we would like to emphasize that,
although
the recent years have been extraordinarily fruitful
for neutrino physics,
yielding important information on the neutrino mixing parameters,
still several fundamental characteristics of neutrinos
are unknown.
Among them,
the Dirac or Majorana nature of neutrinos,
the absolute scale of neutrino masses,
the distinction between the normal and inverted schemes
and
the existence of CP violation in the lepton sector
are very important for our understanding
of the new physics beyond the Standard Model.



\begin{theacknowledgments}
I would like to thank Carlos Pe\~na-Garay for enlightening discussions.
\end{theacknowledgments}

\end{document}